\documentclass[useAMS,usenatbib]{mn2e}

\usepackage{graphicx}
\usepackage{epstopdf}
\usepackage{dcolumn}
\usepackage{bm}
\usepackage{amsmath}
\usepackage{mathrsfs}
\usepackage{amsfonts}
\usepackage[usenames]{color}
\usepackage{epstopdf}

\newcommand{\be} {\begin{equation}}
\newcommand{\ee} {\end{equation}}

\newcommand{\bea}{\begin{eqnarray}}
\newcommand{\eea}{\end{eqnarray}}
\newcommand{\bdm}{\begin{displaymath}}
\newcommand{\edm}{\end{displaymath}}
\newcommand{\ba} {\begin{array}}
\newcommand{\ea} {\end{array}}

\newcommand{\bfg}  {\begin{figure}}
\newcommand{\efg}  {\end{figure}}
\newcommand{\incgr} {\includegraphics}
\newcommand{\mbf}{\mathbf}

\newcommand{\bit}{\begin{itemize}}
\newcommand{\eit}{\end{itemize}}
\newcommand{\fnl}{f_{\rm NL} }

\title[Needlet Bispectrum Asymmetries in the WMAP5]{Needlet Bispectrum Asymmetries in the WMAP 5-year Data}
\author[D.~Pietrobon et al.]
{Davide Pietrobon$^{1,2}$
\thanks{E-mail:davide.pietrobon@roma2.infn.it},
 Paolo Cabella$^{1,4}$,
 Amedeo Balbi$^{1,3}$,
 Robert Crittenden$^2$,
\newauthor  Giancarlo de Gasperis$^1$
and Nicola Vittorio$^1$ \\
\\
$1$ Dipartimento di Fisica, Universit\`a di Roma ``Tor Vergata'', via della Ricerca Scientifica 1, 00133 Roma, Italy \\
$2$ Institute of Cosmology and Gravitation
Dennis Sciama Building
Burnaby Road
Portsmouth, PO1 3FX
United Kingdom\\
$3$ INFN Sezione di Roma ``Tor Vergata'', via della Ricerca
Scientifica 1, 00133 Roma, Italy\\
$4$ Dipartimento di Fisica, Universit\`a La Sapienza, P.~le A.~Moro 2, Roma, Italy\\ }

\begin{document}

\maketitle
\label{firstpage}

\begin{abstract}
We apply the needlet formalism to the Wilkinson Microwave Anisotropy Probe 5-year data, looking for evidence of non-Gaussianity in the bispectrum of the needlet amplitudes.  We confirm earlier findings of an asymmetry in the non-Gaussianity between the northern and southern galactic hemispheres.  We attempt to isolate which scales and geometrical configurations are most anomalous, and find the bispectrum is most significant on large scales and in the more co-linear configurations, and also in the `squeezed' configurations. However, these anomalies do not appear to affect the estimate of the non-linear parameter $\fnl$, and we see no significant difference between its value measured in the two hemispheres.

\end{abstract}


\section{Introduction}
Since the first release of the WMAP satellite data \citep{Bennett:2003bz}, there have been many claims of anomalies in the statistical distribution of CMB temperature fluctuations in the sky (see e.g. \cite{Eriksen:2003db}). For example, there appear to be localised areas which are hotter or colder than would be expected in the concordance $\Lambda$CDM cosmological model with Gaussian statistics (see \cite{Cruz:2004ce}). Also, power seems to be preferentially aligned along a certain direction (dubbed the `axis of evil,'  \cite{Land:2006bn}) and the quadrupole and octopole power appears to be correlated \citep{deOliveiraCosta:2003pu}.  These anomalies were subsequently confirmed with new releases of the
WMAP data \citep{Spergel:2006hy,Nolta:2008ih}. 

Many other studies have highlighted a marked difference in the statistics of the northern and southern galactic skies. 
\cite{Park:2003qd} found an asymmetry in the Minkowski functionals values in the northern and southern galactic hemispheres. 
\cite{Eriksen:2004iu} detected anomalies at large angular scales comparing the amplitudes of temperature power spectra in the
two hemispheres and confirmed the anomalies are present in the n-point correlation function. 
\cite{Vielva2004} studied the kurtosis of Spherical Mexican Hat Wavelets coefficients, discovering a strong
non-Gaussian signal in the southern hemisphere. 
\cite{Hansen:2004mj} reported that the local curvature of the CMB sky exhibited asymmetric behaviour as well. 
\cite{McEwen2008} and \cite{Pietrobon2008AISO} applied two different wavelets constructions to the 5-year
WMAP data, confirming many of these results; they have also been seen
using scaling indices \citep{Rossmanith:2009cy}. \cite{Copi:2006tu} pointed out a lack of power in the north hemisphere in the two point correlation function. 
The presence of these anomalies has been tested against mask effect
and foreground contamination by
\cite{Bernui:2007eu}. \cite{Lew:2008mq} constrains the
direction of the anomaly axis using a generic maximum a posteriori
method. Very recently, \cite{Hansen2008PowAsym} reported that the power asymmetry spans a very large range of angular scales (corresponding to
multipoles $2\le \ell \le 600$): this result is based on an angular power spectrum analysis of the WMAP sky maps. 
A summary of most of these anomalies can be found in \cite{Bernui2008}.

Here, we investigate the CMB anomalies
using the needlets bispectrum\citep{Lan2008needBis} to the WMAP 5-year data; this technique was recently used to constrain primordial non-Gaussianity in the same dataset by \cite{Pietrobon2008NG} and
\cite{Rudjord2009needBis}. For the first time, we analyse the
contribution of different triangle configurations, grouped according to
their size and shape. The paper is organised as follows: In
Sec.~\ref{sec:need_form} we describe the needlet framework; the
data set and the simulations we use are discussed in
Sec.~\ref{sec:results} where we present our results on the north-south
asymmetry and its configuration dependence; finally, in Sec.~\ref{sec:concl} we draw our conclusions.
\section{Needlets Bispectrum Formalism}
\label{sec:need_form}
We perform our analysis of the non-Gaussianity of WMAP 5-year data by means of needlets, which are isotropic wavelets 
with many useful properties \citep{NarcowichPetrushevWard2006,Baldi2006}.  Notably, needlets have bounded support in the harmonic domain, 
while still being quasi-exponentially localised in real space; in addition, they are reasonably straight-forward to implement in practice.  
Thus far, needlets have been successfully applied to the study of the CMB in the context
of the detection of the Integrated Sachs-Wolfe effect (ISW) \citep{Pietrobon2006ISW}, the CMB angular power spectrum estimation
\citep{Fay2008PS}, the study of deviations from statistical isotropy \citep{Pietrobon2008AISO} and the estimation of the primordial
non-Gaussian parameter $\fnl$
\citep{Pietrobon2008NG,Rudjord2009needBis}. Recently the needlet formalism has been extended to polarisation \citep{Geller2008Mat}. We refer to the work by
\cite{Marinucci2008NEE} and \cite{Lan2008needBis} for details on the
construction of a needlet frame and a detailed analysis of its
statistical properties. \cite{Guilloux:2007} provide a similar
construction and discuss an application to component separation in
\cite{Moudden:2004wi}.  A set of needlets has one free parameter, $B$, which controls the width of the filter function; smaller $B$ corresponds to 
a tighter localisation in harmonic space, while a larger value makes it more localised in real space. 
The small correlation among needlets at
different resolutions belonging to the same set can be easily described
analytically and allows for sensitivity when looking for weak signals, like the non-Gaussianity. 
Formally, a needlet function $\psi_{jk}$ is expressed as a quadratic
combination of spherical harmonics which looks like
\be
  \psi_{jk}(\hat\gamma) = \sqrt{\lambda_{jk}}\sum_{\ell}b\Big(\frac{\ell}{B^{j}}\Big)\sum_{m=-\ell}^{\ell}\overline{Y}
_{\ell m}(\hat\gamma)Y_{\ell m}(\xi _{jk})
  \label{eq:needlets_expansion}
\ee
where $\hat\gamma$ is a generic direction in the sky; 
$\xi_{jk}$ and $\lambda_{jk}$ refer to a `cubature point' and a `cubature
weight' respectively which allows the reconstruction on the sphere for the $j$
resolution; the function $b(\ell/B^j)$ is the filter in $\ell$-space.
The needlet coefficients of a field $T(\hat\gamma)$ are given by
the projection of the field itself on the corresponding needlet $\psi_{jk}(\hat\gamma)$:
\be
  \beta _{jk}=\sqrt{\lambda_{jk}}\sum_{\ell}b\big(\frac{\ell}{B^{j}}\big)\sum_{m=-\ell}^{\ell}a_{\ell m}Y_{\ell m}(\xi _{jk})
  \label{eq:needcof}
\ee
In Fig.~\ref{fig:need_coef} we show the needlet coefficients of WMAP 5-year temperature map for $B=2.0$ and $j=4$.  The anomalous bright spots found by \cite{Pietrobon2008AISO} are clearly visible. 
\begin{figure}
\incgr[width=0.6\columnwidth, angle=90]{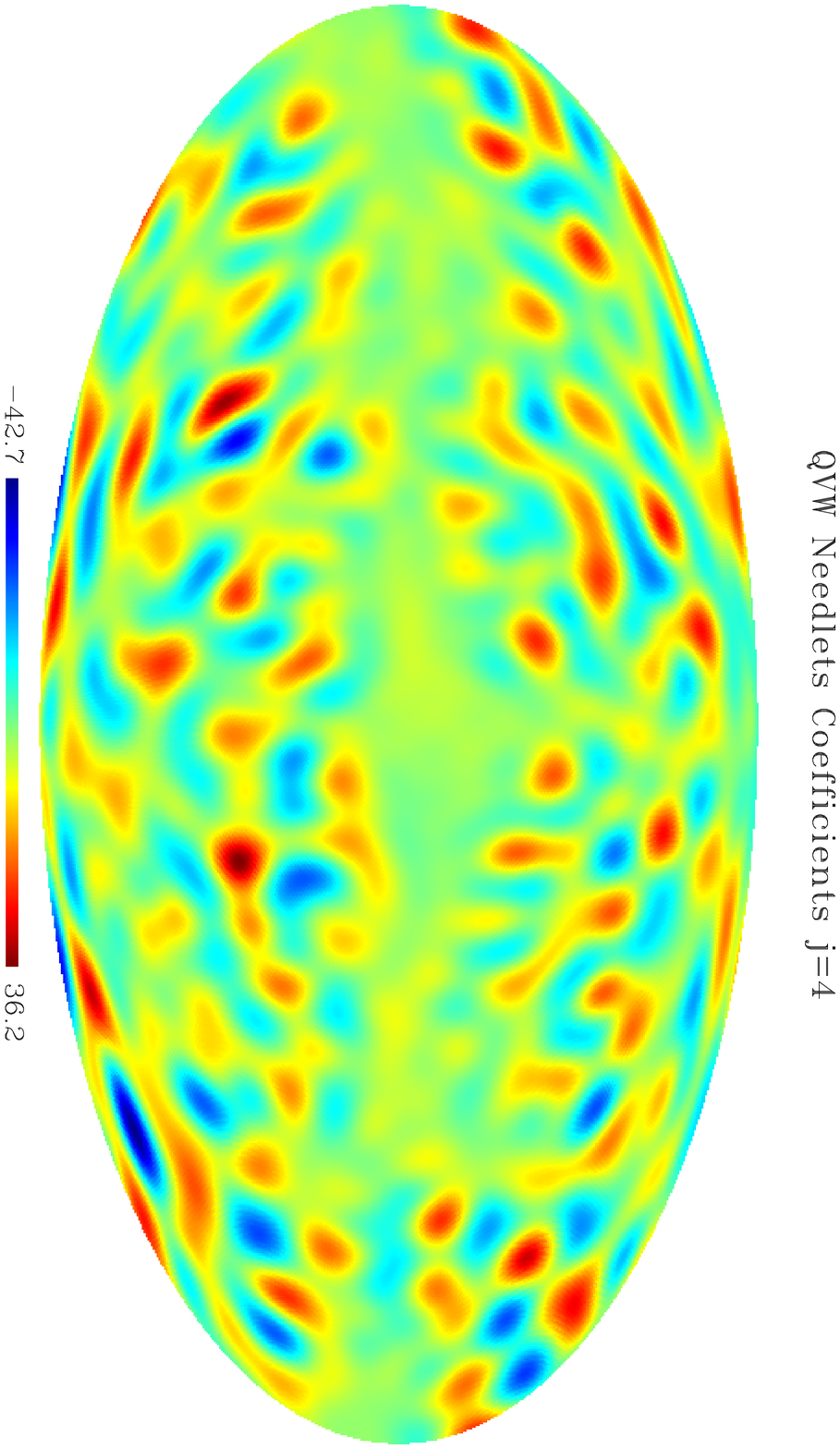}
\caption{Needlet coefficients of the combined Q, V, W map at the resolution $j=4$. The $B$ parameter is fixed to $2$.}
\label{fig:need_coef}
\efg

We next briefly review the
properties of the needlet bispectrum and how it relates to the usual
bispectrum. An extensive discussion is provided in
\cite{Lan2008needBis,Rudjord2009needBis}. See also
\cite{Pietrobon2008NG}. The needlet estimator is defined as follows:
\bea
S_{j_1j_2j_3} &=&\frac{1}{\tilde{N}_p}
\label{eq:needBis}
\sum_k\frac{\beta_{j_1k}\beta_{j_2k}\beta_{j_3k}}{\sigma_{j_1}\sigma_{j_2}\sigma_{j_3}} \\
&\propto& \sum_{\ell_1\ell_2\ell_3}b_{\ell_1}^{(j_1)}b_{\ell_2}^{(j_2)}b_{\ell_3}^{(j_3)}\sqrt{\frac{(2\ell_1+1)(2\ell_2+1)(2\ell_3+1)}{4\pi}} \nonumber \\
&\times&\left(
\begin{array}{ccc}
\ell_1 & \ell_2 & \ell_3 \\
0 & 0 & 0 \\
\end{array}
\right)\hat{\rm B}_{\ell_1\ell_2\ell_3} \nonumber
\eea
where $\tilde N_p$ means the pixels outside the mask and $\hat{\rm
B}_{\ell_1\ell_2\ell_3}$ is estimated bispectrum, averaged over $m_i$s. $S_{j_1j_2j_3}$ can be seen as a \emph{binned bispectrum}, a smooth and
combined component of the angular bispectrum. 
We divide these bispectrum measurements into four classes based on their geometries:  equilateral (\emph{equi}) configurations have 
three equal $j$ values, isosceles configurations (\emph{iso}) have two legs equal (e.g. $j_1=j_2\neq j_3$),  while scalene configurations (\emph{scal}) have three different legs.   Finally we also consider \emph{open} configurations, for which $j_1$, $j_2$ and $j_3$ do not form a triangle 
(e.g. $j_1 > j_2+j_3$);  naively these might be expected to be zero, but since each $j$ includes a range of $\ell$ values, these could include 
signals arising from  $\ell_1$, $\ell_2$ and $\ell_3$ which just satisfy the triangle relations.  Thus open configurations correspond to the most co-linear geometries.   This must be kept in mind for all the configurations; e.g., while the equilateral $j$-configurations will be dominated by triangles roughly equilateral in $\ell$, they will also have contributions from other geometries. 
Separating the needlet bispectrum by the above described configurations may provide insight into the physical origin of possible anomalies. 
For instance \cite{Ackerman:2007nb} and \cite{Erickcek:2008sm} suggest early Universe models which could produce a statistically anisotropic CMB sky.

\section{Statistical Analysis and Results for WMAP 5-year Data}
\label{sec:results}

Next we describe the simulations used in our needlet
bispectrum analysis. We start by producing simulated Gaussian CMB
maps taking into account the beam and
the noise properties of each WMAP-5\footnote{http://lambda.gsfc.nasa.gov/product/map/dr3/m\_products.cfm} channel Q, V and W. From these single-channel maps we
construct an optimal map $T(\hat\gamma) = \sum_{ch} T_{ch}(\hat\gamma)
w_{ch}(\hat\gamma)$ (see \cite{Jarosik2007}), where $\hat\gamma$ represents a direction on the sky and
$w_{ch} = n_h(\hat\gamma)/\sigma^2_{ch}/\sum_{ch}w_{ch}$
where $n_h$ is the number of observations of a given pixel
and $\sigma_{ch}$ the nominal sensitivity of the channel \citep{Hinshaw2008WMAP5}. We apply the ``$j_3$-$j_4$''Kq75 combined
mask described by \cite{Pietrobon2008AISO} and degrade the resulting map to the resolution of $N_{\rm }= 256$.
We extract the needlet coefficients $\beta_{jk}$ from the
simulated maps for a given $B$ and compute  the needlet bispectrum of the reconstructed coefficient
maps according to Eq.~\ref{eq:needBis}. Finally, we calculate $S_{j_1j_2j_3}$ from the real data of the
foreground-reduced WMAP 5-year Q, V and W channels data, using the
same procedure applied to the simulated maps.
To test the Gaussianity of WMAP 5-year data, we
compare the distribution of the $\chi^2=XC^{-1}X^T$ of the simulated dataset to the
value obtained from data, where $X$ is the array consisting of the needlet bispectrum values
calculated via Eq.~\ref{eq:needBis}.
We consider the needlet bispectrum values (indicated by ``\emph{all}'' in the
tables)' and, to identify where the anomalies are concentrated, we split
the analysis in different branches according to the geometry of the
triangles. 
For the chosen $B=2.0$, we have 115 which satisfy the
requirements: 9 equilateral, 56 isosceles, 50 scalene. We define the
remaining 50 configurations as open: we would expect them to be
vanishing except for those which combine multipoles which fulfill the
Wigner selection rules. The correspondence between each needlet scale $j$ and its multipole range is shown in Table~\ref{tab:ell}. 
\begin{table}
\begin{center}
\begin{tabular}{|c|c|c|c|c|c|}
\hline
\multicolumn{6}{|c|}{Large Scales} \\
\hline
  $\mbf{j}$ & 1 & 2 & 3 & 4 & 5 \\
  $\mbf{[\ell_1,\ell_2]}$ & $[2,3]$  & $[3,7]$ & $[5,15]$ & $[9,31]$ & $[17,63]$ \\
\hline
\hline
\multicolumn{6}{|c|}{Small Scales} \\
\hline
  $\mbf{j}$ & 6 & 7 & 8 & 9 & \\
  $\mbf{[\ell_1,\ell_2]}$ & $[33,127]$ & $[65,255]$ & $[132,500]$ & $[263,500]$ & -\\
\hline
\end{tabular}
\end{center}
\caption{Correspondence between angular scale and needlet scale for $B=2.0$.}
\label{tab:ell}
\end{table}

On the full CMB sky, the $\chi^2$ of the data is compatible
with the distribution we obtain from 20,000 Gaussian simulations.
When we calculate the $\chi^2$ on the northern and southern hemispheres
separately, we find a significant deviation from
Gaussianity in the southern hemisphere while the northern hemisphere appears Gaussian, 
having a bispectrum generally somewhat smaller than expected 
(see Table~\ref{tab:asym_chi2}). The results are shown in the histogram plots in
Fig.~\ref{fig:QVW_analysis}.
\begin{figure}
\begin{center}
\incgr[width=.45\columnwidth]{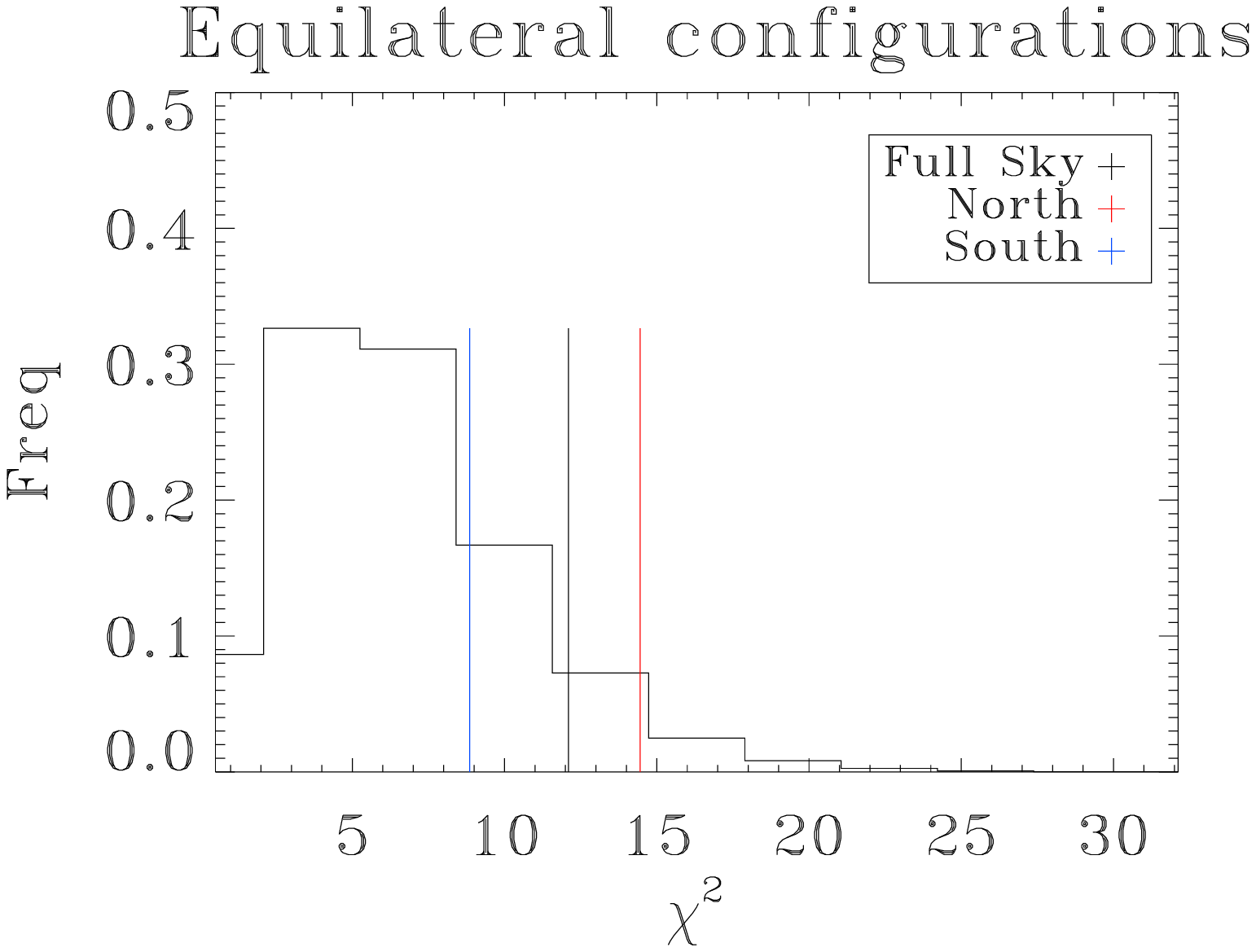}
\incgr[width=.45\columnwidth]{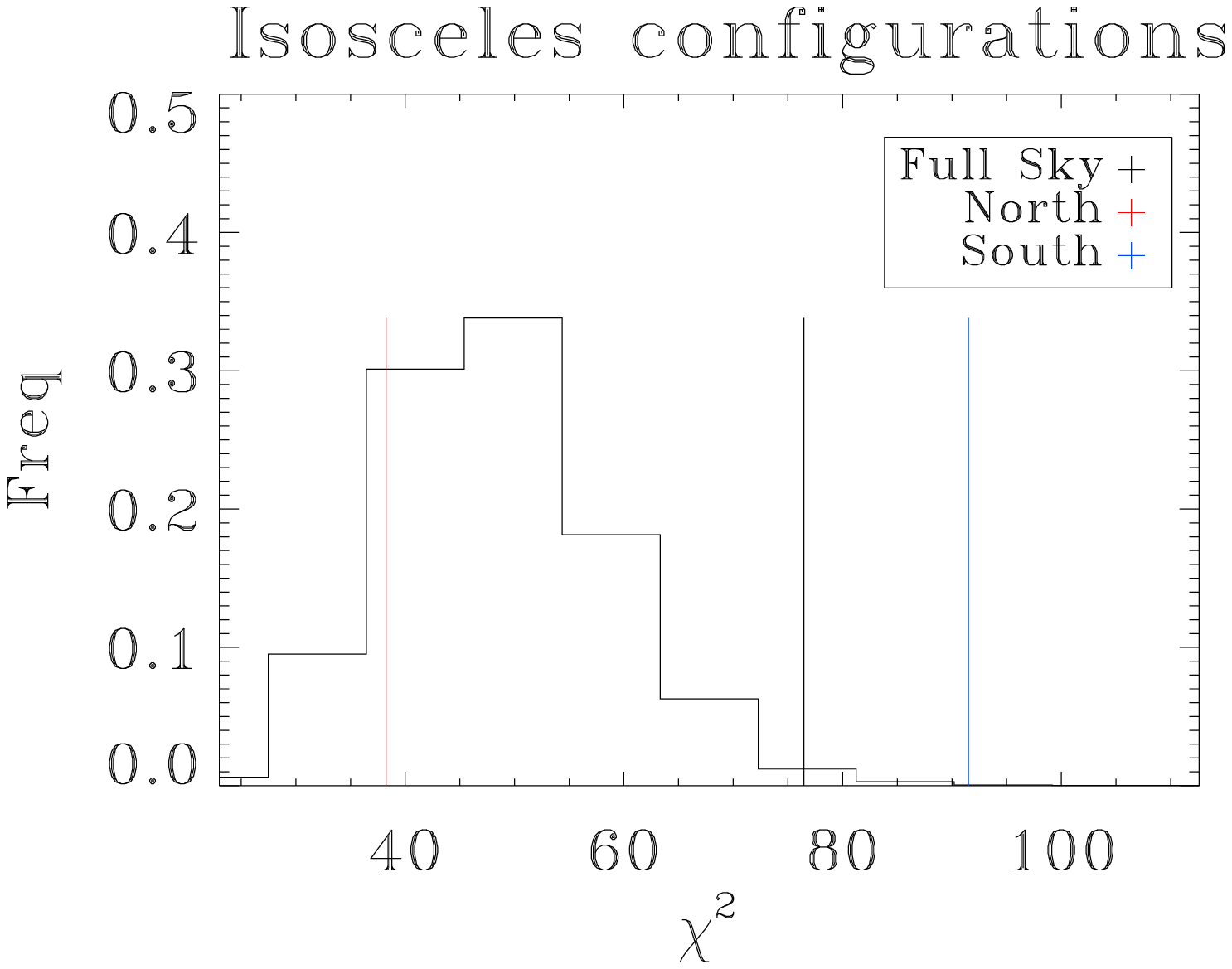}
\incgr[width=.45\columnwidth]{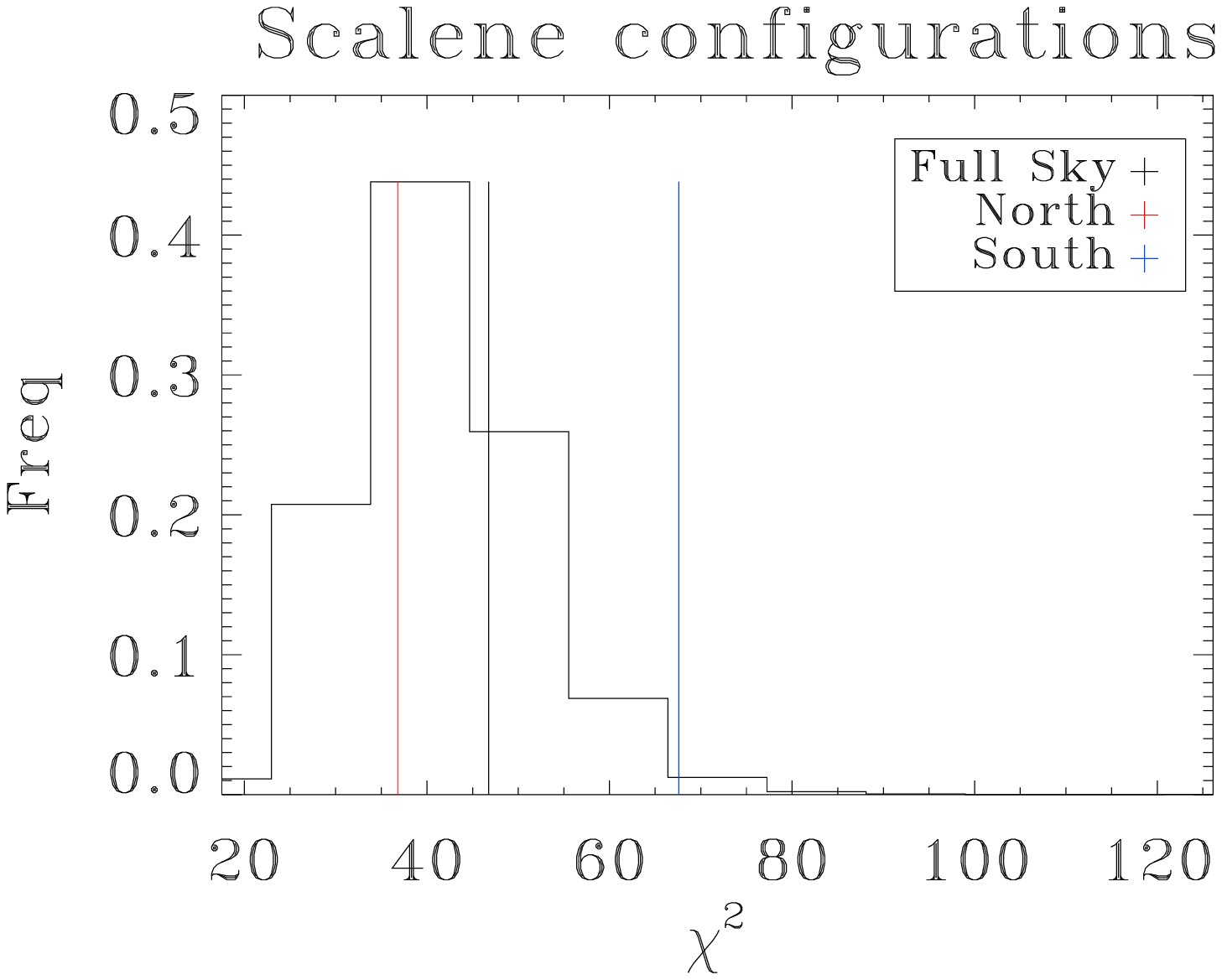}
\incgr[width=.45\columnwidth]{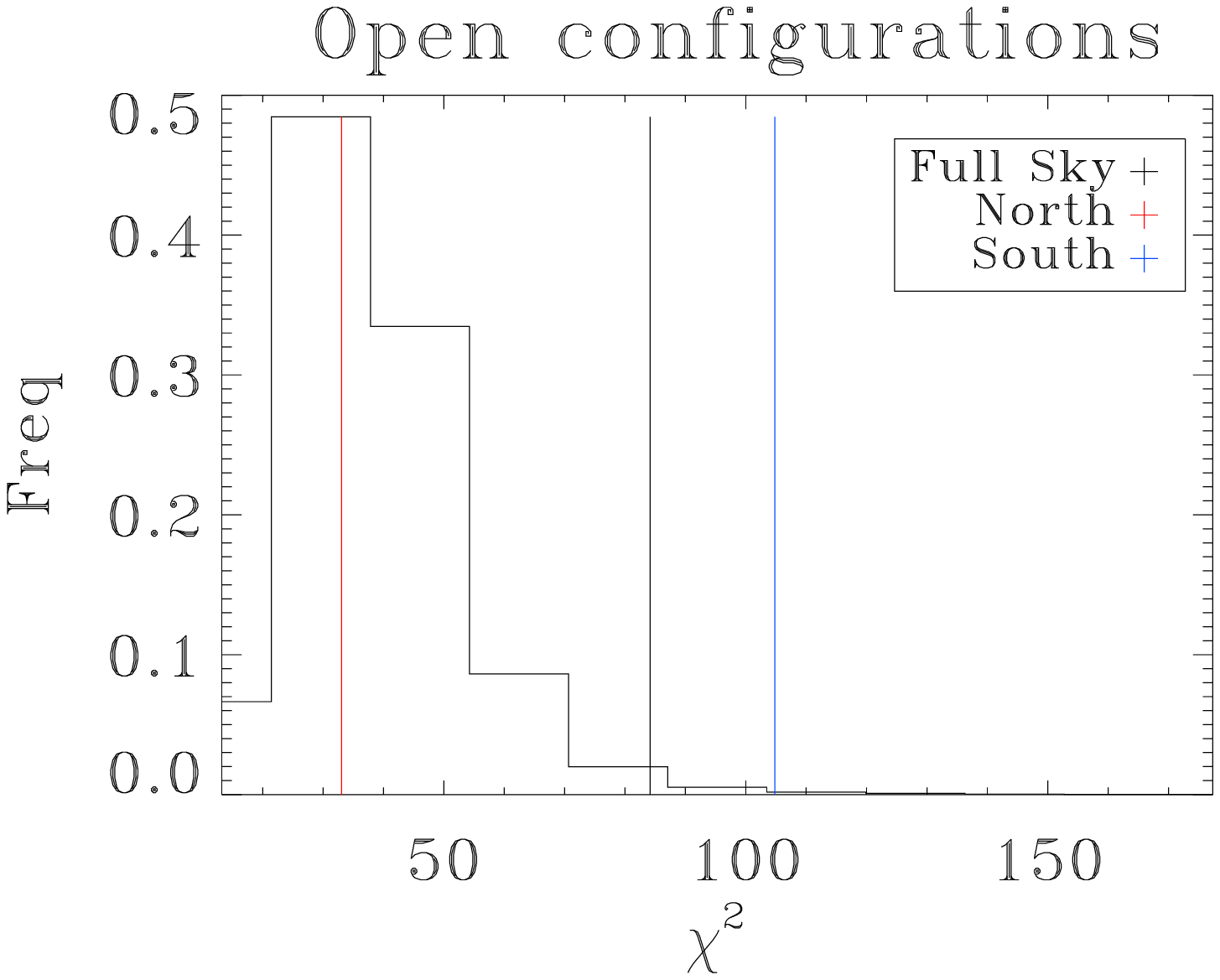}
\caption{Needlet bispectrum $\chi^2$ distribution of the three WMAP 5-year
temperature data. The southern hemisphere is barely compatible with the Gaussian hypothesis, being the blue line which marks the real data $\chi^2$ in the tail of the distribution.}
\label{fig:QVW_analysis}
\end{center}
\end{figure}
Furthermore,
considering the triangle configurations as classified above, we found that
this behaviour is concentrated in all triangle configurations
separately except for the equilateral ones.
The isosceles triangles are perhaps the most interesting
ones since they probe the correlation between the large and the small
angular scales (the so-called `squeezed' configurations), which should reflect a non-local type of
non-Gaussianity. A comparable degree of asymmetry is shown by scalene and
open configurations, which confirm the global lack of power in the
north hemisphere: the points in the northern hemisphere show a lower
scatter. A similar asymmetric behaviour was found by
\cite{Hansen2004,Land2005CubAnom,Eriksen2004} and recently confirmed
by \cite{Hoftuft2009,Hansen2008PowAsym,Groeneboom2009}.
In our analysis we already mask the big anomalous features
present in the southern hemisphere, responsible for about 50\% of the
power asymmetry in the angular power spectrum \citep{Pietrobon2008AISO}.
The results are summed in Table~\ref{tab:asym_chi2}.
\begin{table}
\begin{center}
\begin{tabular}{|c|c|c|c|}
\hline
{\bf conf.}  & {\bf FULL SKY} & {\bf NORTH} & {\bf SOUTH} \\
\hline
\hline
all (115) & $29\%$ & $\mbf{96\%}$ & $\mbf{2\%}$ \\
\hline
equi (9)  & $20\%$ & $11\%$ & $45\%$ \\
\hline
iso (56) & $5\%$ & $\mbf{96\%}$ & $\mbf{0.5\%}$ \\
\hline
scal (50) & $60\%$& $\mbf{90\%}$ & $\mbf{7\%}$ \\
\hline
open (50) & $3\%$& $\mbf{85\%}$ & $\mbf{2\%}$ \\
\hline
\end{tabular}
\caption{Percentage of the simulations with a $\chi^2$ larger than WMAP 5-year data
 for the different triangular configurations of the needlets bispectrum.
 An asymmetry is present in each triangle configuration except for the equilateral, 
and is significant when all the configurations are combined.}
\label{tab:asym_chi2}
\end{center}
\end{table}
Note that equilateral
configurations are directly related to the skewness of the needlet coefficient distributions: the fact that on the whole sky we do not find a strong
deviation from Gaussianity is in agreement with the previous
literature, where only the kurtosis of the distributions showed an anomalous behavior (see \cite{Vielva2004}).

\paragraph*{Large-Small Scale Analysis.} Going more deeply, we focused on small and large angular scales
separately. In particular, with the choice $B=2$, we define the
subset of needlets $j=1$ to $j=5$ as large scale, corresponding
roughly to scales larger than 1 degree; while the subset
$j=6$ to $j=9$ corresponds to the sub-degree scales (see Table~\ref{tab:ell}). We then perform
the same analysis we carried out on the whole needlet set. The results are
shown in Table~\ref{tab:large_small_sets}. The isosceles configurations still show a large
difference between the two hemispheres but the significance is lower
than the whole set analysis. The open configuration result is still
anomalous.
\begin{table}
\begin{center}
\begin{tabular}{|c|c|c|c|}
\hline
{\bf conf.} & \multicolumn{3}{c|}{\bf Large Scales ($j\leq5$)} \\
 & {\bf FULL SKY} & {\bf NORTH} & {\bf SOUTH} \\
\hline
\hline
all (28) & $61\%$& $93\%$ & $14\%$ \\
\hline
equi (5)  & $86\%$ & $26\%$ & $45\%$ \\
\hline
iso (16) & $70\%$ & $90\%$ & $22\%$ \\
\hline
scal (7) & $37\%$& $62\%$ & $15\%$ \\
\hline
open (7) & $3\%$& $38\%$ & $2\%$ \\
\hline
\hline
{\bf conf.} & \multicolumn{3}{c|}{\bf Small Scales ($j\ge6$)} \\
 & {\bf FULL SKY} & {\bf NORTH} & {\bf SOUTH} \\
\hline
\hline
all (20) & $11\%$& $60\%$ & $21\%$ \\
\hline
equi (4)  & $4\%$ & $10\%$ & $36\%$ \\
\hline
iso (12) & $5\%$ & $63\%$ & $8\%$ \\
\hline
scal (4) & $64\%$& $61\%$ & $48\%$ \\
\hline
\end{tabular}
\end{center}
\caption{$\chi^2$ for the WMAP 5-year QVW data compared to
simulations. Top panel large scale study; bottom panel small scale one.}
\label{tab:large_small_sets}
\end{table}
No open configurations exist for the small scale subset $6 \leq j \leq 9$; however for the large scales
 these co-linear configurations are most significantly non-zero
for the biggest contribution of the power. 
For the sub-degree set we did not find an high degree of anomaly, as
summarised in Tab.~\ref{tab:large_small_sets}, though the isosceles
configurations are still significantly different between the two
hemispheres.
Dividing the analysis between the two sets at large and small scales,
we miss the important contribution given by the correlation between
the two, which is indeed crucial for the squeezed 
triangles. We then consider two more sets: one formed by triangles
with one side $j_1\leq5$ and two sides belonging to the small scale
set ($j_2$,$j_3\ge6$). We label this set as LSS and are predominantly squeezed. The second set,
labelled as LLS, is formed by triangles which have $j_1$,$j_2\leq5$ and
$j_3\ge6$ and are predominantly co-linear. We report the results in Tab.~\ref{tab:LSS_set}.
\begin{table}
\begin{center}
\begin{tabular}{|c|c|c|c|}
\hline
{\bf conf.} & \multicolumn{3}{c|}{\bf Correlation (L-S-S)} \\
 & {\bf FULL SKY} & {\bf NORTH} & {\bf SOUTH} \\
\hline
\hline
iso (20) & $23\%$ & $78\%$ & $0.4\%$ \\
\hline
scal (26) & $76\%$& $40\%$ & $51\%$ \\
\hline
open (4) & $32\%$& $35\%$ & $54\%$ \\
\hline
\hline
{\bf conf.} & \multicolumn{3}{c|}{\bf Correlation (L-L-S)} \\
 & {\bf FULL SKY} & {\bf NORTH} & {\bf SOUTH} \\
\hline
\hline
iso (8) & $47\%$ & $94\%$ & $20\%$ \\
\hline
scal (13) & $62\%$& $98\%$ & $15\%$ \\
\hline
open (39) & $3\%$& $88\%$ & $2\%$ \\
\hline
\end{tabular}
\end{center}
\caption{$\chi^2$ for the WMAP 5-year QVW data compared to
simulations: correlation large-small scale. Top panel LSS set; bottom panel LLS set.}
\label{tab:LSS_set}
\end{table}
The isosceles triangles belonging to the LSS set are very
anomalous in the south hemisphere. The LLS set in characterised by an
anomaly in the open configurations. Interestingly the lack of signal in
the northern hemisphere is evident in the LLS, while the LSS distribution appears more typical.

\paragraph*{Further analysis.} Since the first anomalies were found, several methods have been
applied to search for a specific direction in the sky which maximises
the asymmetry. Indeed, many works identify a direction very close
to the ecliptic poles
\citep{Hansen2004,Land2005CubAnom,Raeth:2009,Hansen2008PowAsym}. In
particular a dipole modulation has been proposed as a possible
explanation for such pattern in \citep{Hoftuft2009}. In order to see whether a similar
modulation underlies the asymmetry we detect, we
rotated the reference frame, spanning uniformly the sky,
and recomputed our statistics for the WMAP data with the new
north-south definition. Since the set of simulations we used assume
isotropy and homogeneity, the rotation of the reference frame does not
affect their statistics and the covariance matrix we applied in our
previous analysis.

The result is shown in Fig.~\ref{fig:modulation}, where we
plotted the reduced $\chi^2$ for the southern hemisphere in the
particular case of the isosceles configurations, as a function of the
north pole direction. We chose the isosceles triangles since they show the
highest degree of asymmetry. A pattern is clearly
visible, but the direction which maximises the anomaly seems to be
orthogonal to the one reported by other
authors with different estimators. This may suggest that either the
3-point correlation function couples differently to the dipole
modulation, or the nature of the asymmetry we measured is different. Whether this direction depends upon the shape and
the angular scale we consider, its significance and the link with the
direction found in literature are interesting
questions, which require a dedicated study and they will be
addressed in a following paper.
\bfg
\center
\incgr[width=.6\columnwidth,angle=90]{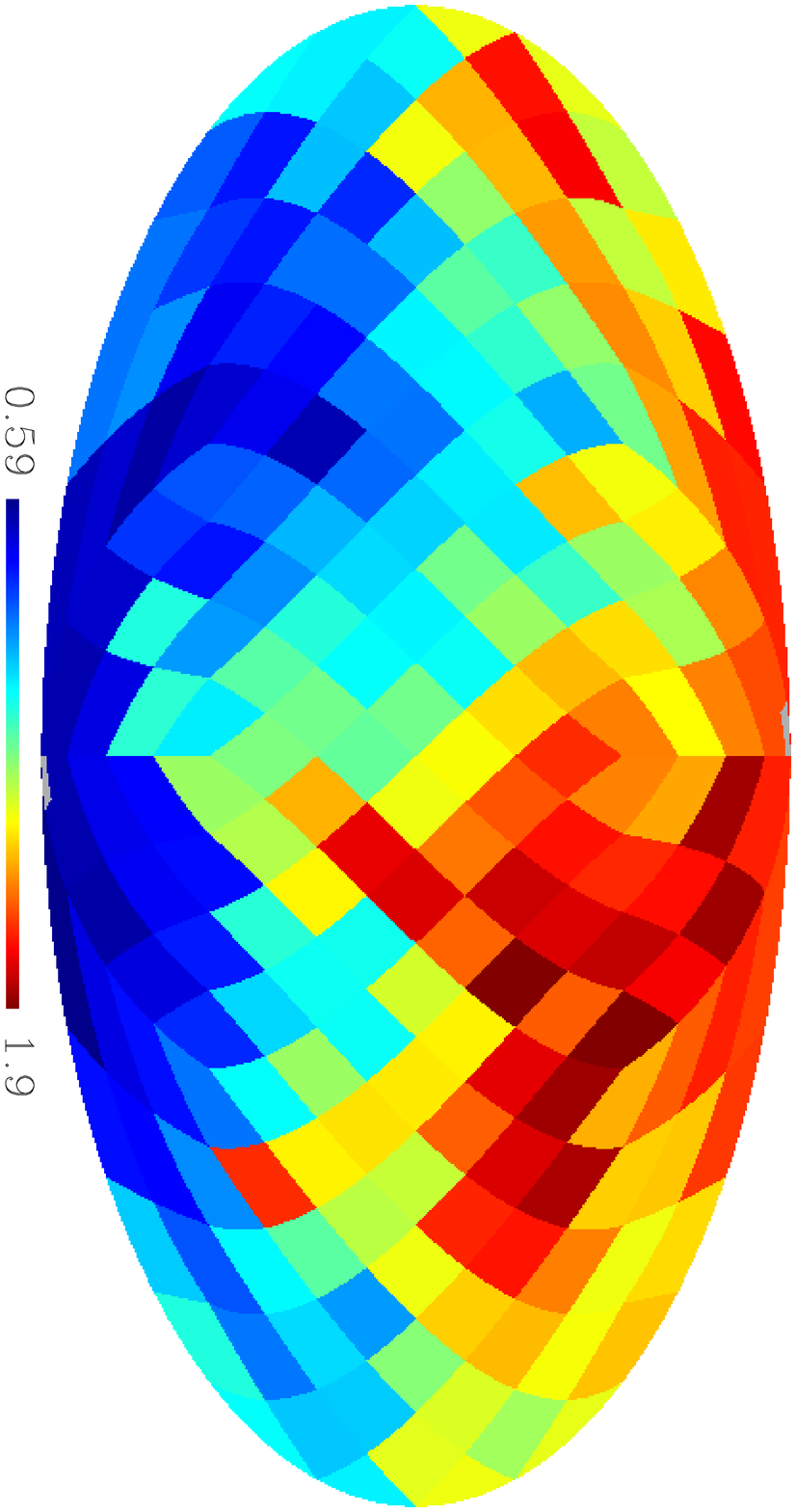}
\caption{Reduced $\chi^2$ of the southern hemisphere as a  function of
  the north pole definition in the most anomalous case of the
isosceles configurations. The gray squares mark the standard z-axis.}
\label{fig:modulation}
\efg

Finally, we checked whether the sky
asymmetries we detected affect the measure of the primordial non-Gaussianity
parameter. A complete review on the nature of this parameter may be found in
\cite{Bartolo2004NGreview} and \cite{Fergusson:2008}; recent constraints from
CMB experiments can be found in
\cite{Smith2009fnlFore,Curto2008Archeops,WMAP5Komatsu2008,DeTroia2007b2k}
 while \cite{Slosar:2008} constrained $\fnl$ through the galaxy
distribution. Limits on $\fnl$ using wavelets are discussed in
\cite{Curto2009NG,Cabella2004,Mukherjee2004}.
We estimate $\fnl$ performing the same analysis described in
\cite{Pietrobon2008NG}, applying the estimator
\begin{equation}
\fnl = \frac{X^{d\,T} C^{-1} X^{\rm th}}{X^{\rm th \,T} C^{-1} X^{\rm th}}
\end{equation}
to the WMAP 5-year dataset. Here X is a vector composed by the values of needlets bispectrum for a
given triangular configuration according to Eq.\ref{eq:needBis}. The
covariance matrix $C$ is calculated from 20.000 Gaussian simulations, since its dependency on
 $\fnl$ is negligible (e.g. see \cite{SpergelGoldberg1999}).
The theoretical non-Gaussian template $X^{th}$ was calculated via Monte
Carlo simulations over the 100 primordial non-Gaussian maps \citep{Liguori2007NGMaps}.
 Since we know the CMB sky is asymmetric, showing more non-Gaussianity in the  southern hemisphere, we carried out 
a split north-south analysis to see if the asymmetry extends to differences in the
$\fnl$ estimate.
 Recently, \cite{Curto2009NG} and \cite{Rudjord:2009} targeted the same
issue
, finding no evidence of
$\fnl$ varying on the sky.
 We do not find a significant deviation between the 
values measured in the two hemispheres, while the error
bars become significantly larger due to the reduced number of pixels
used to calculate the needlet bispectrum.
\section{Conclusions}
\label{sec:concl}
In this paper we used the needlets bispectrum to investigate the
presence of anomalies in the WMAP 5-year data. For the first time
we exploited the bispectrum formalism analysing the triangle
configurations according to their shape. By splitting the $\chi^2$ analysis of the needlets 
bispectrum for the northern and southern hemispheres we found that the southern sky is barely
compatible with the Gaussian hypothesis while the northern hemisphere is
characterised by a lack of global bispectrum signal. This is complementary to what
found by applying different statistics: power spectra
\citep{Hansen2008PowAsym}, bispectrum \citep{Land2005CubAnom} and
n-point correlation functions \citep{Eriksen:2004iu}.
We distinguished equilateral,
isosceles, scalene and open configurations and compared the power
present in the data to random Gaussian simulations.
The most anomalous signals in the southern Galactic hemisphere arise in the squeezed configurations 
(isosceles, large-small-small) and in the very co-linear configurations (open, large-large-small).  
This kind of information should be useful as a means to find out the physical origin of the anomalies.
While the large squeezed signal hints at a local type of non-Gaussianity, this is not bourne out when 
an optimal estimator tuned specifically to this type of non-Gaussianity is used.   
We investigated the effect of hemispherical asymmetry on the
measurement of $\fnl$ finding no significant discrepancy between north
and south.
 As consistency check, we performed the same tests (anomalies and $\fnl$
estimates) with a different needlets parameter ($B=3.5$) and for the channels
Q,V and W separately and found consistent results. 
\section*{Acknowledgements}
We thank Frode K.~Hansen, Michele Liguori and Sabino Matarrese for
providing us with the primordial non-Gaussian map dataset. We are
grateful to Domenico Marinucci and Marcella Veneziani for useful
discussions. The ASI contract LFI activity of Phase2 is acknowledged.

\bibliography{Bnsasym}

\end{document}